\documentclass[preprint2]{emulateapj}
\usepackage{graphicx}
\usepackage{natbib}
\usepackage{epsf}

\def\solmas{$\mathrm{M_\odot}$~}
\def\solmasp{$\mathrm{M_\odot}$}
\def\tcr{$t_{\rm cr}$}
\def\tff{$t_{\rm ff}$}

\def\rhooph{$\rho$ Oph}

\newcommand{\mHt}{{\rm H_{2}}}
\def\simless{\mathbin{\lower 3pt\hbox
   {$\rlap{\raise 5pt\hbox{$\char'074$}}\mathchar"7218$}}}
\def\simgreat{\mathbin{\lower 3pt\hbox
   {$\rlap{\raise 5pt\hbox{$\char'076$}}\mathchar"7218$}}}
\newcommand{\zfive}{${\rm Z} = 10^{-5} \: {\rm Z_{\odot}} $}

\newcommand{\rhovol}{ \langle \rho \rangle _{\rm vol} }

\begin{document}

\title{On the constancy of the high-mass slope of the initial mass function}

\author
{Paul C.\ Clark$^1$,  Simon C.\ O.\ Glover$^1$,  Ian A.\ Bonnell$^2$ \&  Ralf S.\ Klessen$^1$}

\affil{$^{1}$Zentrum f\"ur Astronomie der Universit\"at Heidelberg, Institut
f\"ur Theoretische  Astrophysik, Albert-Ueberle-Str.\ 2, 69120 Heidelberg, Germany
\break email: pcc@ita.uni-heidelberg.de, sglover@ita.uni-heidelberg.de, rklessen@ita.uni-heidelberg.de \\
$^2$School of Physics \& Astronomy, University of St Andrews, North Haugh, St Andrews, Fife, KY16 9SS, UK \break email: iab1@st-and.ac.uk}

\begin{abstract}
The observed slope at the high-mass end of the initial mass function (IMF) displays a remarkable universality in a wide variety of physical environments. We predict that competitive accretion, the ongoing accretion of gas from a common reservoir by a collection of protostellar cores, can provide a natural explanation for such a universal slope in star forming regions with metallicities ${\rm Z} \ga 10^{-5} \ {\rm Z_{\odot}}$. In our discussion, we point out that competitive accretion will occur whenever a gaseous region has multiple Jeans masses of material and contains large-scale motions that are controlled by the gravitational potential well. We describe how and when these conditions can be reached during the chemical enrichment of the Universe, showing that they can occur for a wide range of metallicities and environmental conditions. We also discuss the ability of other physical processes to limit the effects of further accretion onto protostellar cores. Current theoretical and numerical studies show that competitive accretion is robust against disrupting effects -- such as feedback from young stars, supersonic turbulence and magnetic fields -- in all but the most extreme cases. 

\end{abstract}

\maketitle

\section{Introduction}
\label{sec:intro}

The origin of the stellar initial mass function (IMF) remains one of the central topics of research in star formation theory and observation. The IMF describes the distribution of masses with which stars are born, and typically takes a power-law form at the high mass end ($\ga 0.3$ \solmasp) of $dN \propto m^{-\alpha} dm$, with $\alpha$ taking a value of around $-2.35$ \citep{Salpeter1955}. The distribution of masses then turns over at lower masses, and is typically fitted with either a series of broken power-laws (e.g. \citealt{Kroupa2001, Kroupa2002}), or with a log-normal function \citep{Chabrier2003}.

While there may be some observational debate regarding details around the peak and turnover in the IMF (e.g. \citealt{Elmegreenetal2008}), the high-mass end is typically well described by the Salpeter slope, for a wide variety of environmental conditions. For example, Orion \citep{HillenbrandCarpenter2000} and \rhooph~\citep{LuhmanRieke1998} both exhibit Salpeter-like IMFs above $\sim$ 0.5 \solmasp, despite the fact that the Orion Nebula cluster contains a quartet of ionizing sources at its center, while \rhooph~is a relatively quiescent region of star formation. Even in more extreme evironments, such as R138 in Doradus \citep{MasseyHunter1998} -- which also lies outside our own Galaxy in the somewhat lower metallicity Large Magellanic Cloud -- the high-mass end of the IMF also appears to follow a Salpeter slope. Only for the globular clusters, the relics of extremely low-metallicity star formation, does the description of the high-mass end of the IMF start to become less clear, since the oldest clusters will have lost all stars $\ga 0.8$ \solmas due to stellar evolution. In this case, we must rely on more indirect measurements of the IMF \citep{ParescedeMarchi2000, deMarchietal2000, Chabrier2003, deGrijsParmentier2007}, with some studies suggesting a somewhat flatter than Salpeter slope for masses $\ga 0.5$ \solmasp \citep{deMarchietal1999, Kochetal2004, deMarchiPulone2007}. 

The Salpeter slope is significant for two reasons. First, the fact that the power is negative shows that low-mass stars are significantly more abundant than their high mass counterparts. Second, the fact that the slope is steeper than -2 shows that most of the mass from the star formation process goes into the low-mass stars. Our current theoretical understanding of the first stars in the Universe -- the so-called Population III (or Pop. III) stars -- suggest that this was not always the case, as these first stars are expected to have had very large masses (see e.g.\ \citealt{BrommLarson2004,Glover2005}). However, solar-mass stars with metallicities as low as [Fe/H] = $-5.9$ have been observed in the Galactic halo \citep{Frebeletal2005, Frebeletal2008}, and although they cannot by themselves constrain the full IMF at these very low metallicities, their existence is suggestive of an early transition from an IMF dominated by massive stars to one dominated by low-mass stars.

For present-day star formation, there are two contrasting theories for the origin of the IMF. The first suggests that the vast majority of the final stellar mass comes from a local reservoir which envelopes the protostar, and is typically referred to as `isolated' star formation \citep*{Shuetal2004} or monolithic collapse \citep{MckeeTan2003}. The mass contained in this reservoir is set by some process that breaks the cloud up into dense structures, such as gravitational fragmentation (e.g.\ \citealt{Larson1973}), turbulent fragmentation (e.g.\ \citealt{PN2002, HennebelleChabrier2008}), or a combination of both \citep{KlessenBurkert2000, Klessenetal2000, KlessenBurkert2001}. These isolated reservoirs then feed their protostar (or small $N$ group) without further influence from the surrounding cloud. 

Competitive accretion \citep{Zinnecker1982,Bonnelletal2001a, Bonnelletal2001b} presents an alternative picture, in which the high-mass slope of the IMF is the result of continued accretion over and above the initial local collapse of the pre-stellar core. This model makes use of the fact that the majority of stars form in clusters \citep{Ladas2003}, with the protocluster `clump' providing the reservoir for on-going accretion. The dynamics of the gas and protostars, that naturally arise due to the gravitational potential of the protocluster core, provide a distribution in the accretion rates that results in a distribution of protostellar masses. Stars which achieve masses larger than the typical Jeans mass in the cloud, which are those following the Salpeter-like section of the IMF, do so by accreting gas from a region which is larger than the average Jeans length \citep*{BVB2004}.

In previous studies of star formation occurring in low metallicity gas at high redshift, attention has focussed primarily on the isolated star formation model. Typically, these studies have sought to determine the characteristic mass scale for fragmentation of the gas through the consideration of its chemical and thermal properties, either within the framework of a simplified dynamical model (e.g. \citealt{Omukai2000,Omukaietal2005}) or within a full cosmological hydrodynamical simulation 
(e.g. \citealt{SmithSigurdsson2007,Smithetal2009}). How this characteristic mass goes on to describe the type  of IMF that we see today, one which covers a wide range in mass, has yet to be addressed.

In this paper, we focus our attention on the competitive accretion model for the IMF, suggesting that it can provide a simple mechanism for a universal high-mass slope, over a wide range in metallicities. First, we give an overview of competitive accretion in \S \ref{sec:compacc}, stressing that it is fairly insensitive to the conditions in the young cluster (such as initial density or velocity profiles). We also discuss the conditions required in order for competitive accretion to dominate over the isolated fragmentation models. In \S\ref{sec:ob}, we briefly discuss the observational evidence from local star-forming regions that favours the competitive accretion model. Based on the expected thermodynamic behavior of the gas, we show in \S \ref{sec:lowz} that the conditions for competitive accretion can be achieved readily in low-metallicity gas, suggesting that a Salpeter-like slope at the high-mass end of the IMF is a universal outcome of clustered star formation. In \S \ref{sec:prev} we highlight a number of processes which could lessen the effects of competitive accretion, but argue that it is unlikely that these would be able to prevent the phenomenon entirely, provided star formation occurs in bound clusters.

\section{An overview of competitive accretion}
\label{sec:compacc}

Competitive accretion is the process whereby individual stars, contained in a bound group or cluster, compete to accrete from a communal reservoir of gas. This process requires that there exists a significant, distributed component of gas and that the gas and stars are mutually bound in the overall gravitational potential. Lastly, the gas needs to have some freedom to move in response to the local and global potential of the system. This last is unavoidable in a bound system where star formation occurs and any impediment such as turbulence or magnetic fields can only act to slow the accretion. (A discussion of such processes is given in \S \ref{sec:prev}).

In order for competitive accretion to generate an IMF, the accretion rates must necessarily vary considerably from star to star. This variation in the accretion rate occurs in the picture outlined above, since the accretion rate depends on a combination of the star's mass, the local gas density, the relative velocity of the gas and the star, and the position of the star in the gravitational potential of the full system. Bonnell et al.~(2001a) pointed out that there are two regimes of accretion, depending on the relative gas-star velocity and the tidal field of the full cluster potential (i.e.\ that from the sum of the gas {\em and} stars). In the regime where the gas and stars are following similar trajectories, due to motion dictated by the cluster potential, then the cluster tidal field is the limiting effect and determines the accretion rate. This tends to be the case when protostars and the surrounding gas are both infalling toward the cluster center, which typically occurs during the fragmentation of the gas, and thus the formation of new stars (e.g. \citealt{Bonnelletal2008}).  If the stars are virialised or otherwise have velocities uncorrelated to the gas, then the ability of a star to capture passing gas, known as Bondi-Hoyle accretion, dominates the accretion process. Finally, tidal accretion also describes the mass evolution of the massive stars located at the centre of the cluster, as they are more or less static and feed off the gas that gets funneled to the center of the cluster, as it falls down the potential well.

A forming protocluster is envisioned to start out in the tidal accretion regime, with newly formed fragments and gas having low relative velocities. The protostellar velocities may also be low relative to the centre of mass of the system at this point. The accretion rate is a measure of a star's ability to attract gas to itself. The stars need not move through the system, sweeping up large volumes of gas,  in order to have significant accretion rates. Instead, the gas responds to both the local and cluster potential and moves accordingly. Bonnell et~al.~(2001b) showed that for clusters with a $\rho\propto r^{-2}$ mass distribution, as is expected under gravitational collapse and as observed in stellar clusters, tidal accretion results in a fairly shallow IMF, with $dN/dm \propto m^{-1.5}$. 

Bondi-Hoyle accretion occurs after the stars have virialised, for example via interactions in the cluster center, generating large velocities relative to that of the infalling gas. A star's ability to accrete is then set by the star's mass and velocity in addition to the gas density
\begin{equation}
\label{equ:bh}
\dot{m} \propto \rho \ \frac{m_*^2}{v_{\rm rel}^3}\ .
\end{equation}
Zinnecker (1982) first showed how Bondi-Hoyle accretion in a uniform density gas generates a Salpeter-like IMF $dN/dm \propto m^{-2}$. For more realistic clusters with a density profile and an initial mass segregation, the resulting IMF is steeper with $dN/dm \propto m^{-2.5}$ (Bonnell et al.~2001b).

Gas falling into a bound system's gravitational potential has to follow the local potential gradient. This gas can only settle in local potential minima, which are created by the individual stars. Even in the cluster centre, the local gradient will be defined by the stars nearby and hence the gas will settle into an individual star's potential well. Gas cannot gather in regions where the gradient is not zero and hence cannot form new cores from unbound gas. Ongoing star formation is limited to infalling cores that are already bound as they enter the cluster potential, as otherwise the tidal fields of the system will shred any unbound gas \citep{Bonnelletal2008}. 
 
There are a number of aspects of competitive accretion which often cause confusion. First, the stars do not need to move in order to participate in competitive accretion, as in this case it is the tidal radii of the stars inside the cluster which determine the accretion rates. Second, most stars in a cluster actually lose the competition and remain relatively low-mass stars (Bonnell \& Bate 2006). Finally, the most massive star, forming in the centre of the cluster, accretes more through  tidal accretion than Bondi-Hoyle accretion. Instead of moving at high relative velocity, the most massive star remains in the centre of the potential -- which it helps to define -- and the gas flows down to it.
 
The resulting IMFs are robust in terms of the cluster properties. A collapsing cloud which is not completely cold always develops a $\rho \propto r^{-2}$ density profile which generates a shallow  IMF for the lower mass stars. The high-mass Salpeter-like IMF depends primarily on the $m_{*}^{2}$  nature of the accretion rate and, as discussed above,  generates an IMF which varies from $\alpha=-2$ to $\alpha=-2.5$, depending on the density profile and any initial mass segregation. 

\subsection{Fragmentation and timescales}
\label{sec:fragtime}
For competitive accretion to occur requires that the cloud is able to promptly fragment \citep{Pringle1989}, and that the gas can accrete on timescales shorter than the gas removal timescale due to feedback, 
\begin{equation}
\label{equ:timescales}
\tau_{\rm form} \le \tau_{\rm acc} \le \tau_{\rm emerge}. 
\end{equation}

\noindent The first condition requires that the timescale of collapse of the individual fragments be shorter than the collapse timescale of the system \citep{Bonnelletal2001a}. This will occur as long as the system contains many Jeans masses of gas and there is significant non-linear structure to act as seeds for the fragmentation. The important timescale here is the free-fall time for the system, \tff, defined by $t_{\rm ff} = (3\pi/32G\rhovol)^{1/2}$, where $\rhovol$ is defined to be the volume-averaged density in the protocluster  \citep{Bonnelletal2001a}.  This is also the timescale for stars to attain virialised velocities as they react to the cluster potential, and thus enter the Bondi-Hoyle accretion regime. Even if the cloud contains turbulent motions, the turbulent crossing time (\tcr) must be comparable to \tff, provided that the cloud is gravitational bound,  since $t_{\rm ff}/t_{\rm cr} = (\pi E_{\rm k})/(\sqrt{10}\, |E_{\rm p}|)$,  where $E_{\rm k}$ and $|E_{\rm p}|$ are (respectively) the kinetic and gravitational eneries, for a uniform density spherical cloud. As such, turbulent motions typically do not alter the accretion within the cluster (e.g. \citealt{Klessen2001, BVB2004, SchmejaKlessen2004}; note also that a more detailed discussion of the role played by turbulence is given \S \ref{sec:prev}). 

The second condition is simply that the cluster of protostars must be able to accrete some useful fraction of the background reservoir before the reservoir is removed via feedback processes. Again,  \citet{Bonnelletal2001b} have shown that this is comparable to the mean free-fall timescale for the cluster as a whole. Naturally, additional processes may `heat' the gas in the cluster's potential well, to the point where \tff~is no longer an accurate description of how the ambient gas is accreted. However, current numerical simulations seem to suggest that this is not the case in young clusters, and a fuller discussion of these processes is given in \S \ref{sec:prev}. The fact that \tff~is the governing timescale for both these conditions is part of the reason why the competitive accretion process is so robust once the general conditions are met, since it is not a strong function of the density.

The fragmentation depends crucially on the thermodynamics of the gas, which determines the number of Jeans masses available \citep{Larson1985, Larson2005, Jappsenetal2005, Clarketal2005}. If the gas remains isothermal such that the cooling balances any heating, then even though the local Jeans mass, $M_J \propto T^{3/2} \rho^{-1/2}$, decreases during collapse, the number of Jeans masses available for fragmentation remains constant. This is a consequence of the non-homologous nature of the collapse, which typically generates a density profile $\rho \propto r^{-2}$, leaving a central core of mass $M_{\rm core} \propto \rho^{-1/2}$ where fragmentation can occur. In the $\rho\propto r^{-2}$ envelope, tidal forces will stop any fragmentation. The major result is that the number of fragments produced is given by the number of Jeans masses available in the initial conditions, with small adjustments due to the local geometry \citep{Larson1985, Bastienetal1991, BurkertBodenheimer1996, KlessenBurkert2000, KlessenBurkert2001}. 

If the initial conditions contain one Jeans mass, then the cloud needs to cool during collapse in order to fragment \citep{ReesOstriker1977, Arcoragietal1991}, whereas even if the system contains many Jeans masses, fragmentation will be increasingly suppressed if the cloud heats up with increasing density. Thus, fragmentation will predominantly occur in or follow from evolutionary regions where the temperature decreases with increasing density. Once fragmentation does occur, if it produces bound systems which can accrete, then competitive accretion will necessarily follow and produce a ubiquitous shape to the IMF.

It has also been demonstrated that competitive accretion is insensitive to the initial mass distribution of the protostars (or protostellar cores) that come out of the fragmentation process (Bonnell et~al.~1997). Motions within the cluster, along with further accretion, tend to re-organize the protostars such that the resulting mass function produced in the cluster is the same. As such, competitive accretion will produce a Salpeter slope at the high-mass end that is robust against the details of how the gas is fragmenting.

In summary, we suggest that any region of collapsing, fragmenting, gas, will undergo a competitive, Bondi-Holye-style accretion process. The small degree of variation in the high-mass IMFs that this process produces is what we believe to be responsible for the near universal IMFs observed in all resolved stellar populations. In the following section we review the conditions in local star-forming regions, showing how they exhibit the properties that have been discussed here.  In \S \ref{sec:lowz}, we go on to discuss how these competitive-accretion-friendly conditions can be achieved for a wide range of metallicities.

\section{Observational picture from star-forming regions}
\label{sec:ob}

Given the picture outlined in \S \ref{sec:compacc}, it is clear that for competitive accretion to be important in real star-forming regions, they must contain multiple Jeans masses and contain gas whose properties -- such as the primary fragmentation mode and motions -- are described by timescales that are shorter than the time for the cluster to emerge from the cloud. Indeed, proto-clusters that are globally collapsing would provide a particularly fertile environment for competitive accretion.

The results from recent observational surveys of nearby star-forming regions  favor this picture \citep{WilkingLada1983, Motteetal1998, TestiSargent1998, Johnstoneetal2000, Johnstoneetal2001, Johnstoneetal2006, NutterWardThompson2007, Andreetal2007}. Typically, the clouds contain upwards of 500 \solmas and the temperatures and densities are in the range of 10--20 K and $10^{4}$--$10^{5}$ cm$^{-3}$, respectively. Such low temperatures are to be expected since the background interstellar UV is typically highly attenuated in regions of star formation due to their high column density (e.g.\ \citealt{StamatellosWhitworth2003, Youngetal2004, Shirleyetal2005}). The observed Jeans mass in the clouds is therefore $\la 2$\solmasp, and the corresponding Jeans length is $\la 10,000$ AU. Not only do these clouds therefore contain multiple Jeans masses, but the distance between pre-stellar cores is also observed to be similar to the Jeans length in the regions of active star formation. This suggests that the clouds are free to undergo an initial Jeans-like fragmentation. The free-fall time of the gas in these star forming regions is also quite short, of the order $10^{5}$ yr or less, much shorter than the 4 - 5 million years it takes for the proto-stellar clusters to throw off their natal envelopes \citep{Leisawizetal1989}.

There is also evidence that these very young regions are not only multiply Jeans unstable, but are also kinetically very cold. \citet{Andreetal2007} show that the virial mass derived from the N$_{2}$H$^{+}$ line emission in the L1688 cloud in \rhooph~is an order of magnitude smaller than the value derived from the C$^{18}$O observations \citep{WilkingLada1983}. Similar results have been found for other young regions such as NGC 2264 \citep{Perettoetal2006, Perettoetal2007} and NGC 1333 \citep{Walshetal2006, Walshetal2007}. Such kinetically cold initial conditions are encouraging for the competitive accretion scenario, since they suggest that the observed pre-stellar cores will be brought together by the large-scale collapse of the region. 

\section{The dependency on metallicity}
\label{sec:lowz}

Although the observations described in \S\ref{sec:ob} suggest that  the gas in local star-forming regions is multiply Jeans unstable and hence in an ideal state for competitive accretion to occur, we have no such direct probe of the state of the star-forming gas in lower metallicity systems. However, from a theoretical viewpoint, it does not seem difficult to arrange for the gas to achieve the conditions required for competitive accretion. For competitive accretion to occur, we require multiple Jeans masses of
gas, plus enough density substructure to act as seeds for fragmentation. If we start with $\sim 1$
Jeans mass of gas, then, as explained in \S\ref{sec:fragtime}, in order to produce multiple Jeans mass, 
the gas temperature must decrease during the collapse, i.e.\ the gas must cool. Moreover, if we are to preserve
our density substructure for long enough for it to cause fragmentation, then we must be able to cool
quickly, within less than a dynamical time, since we do not expect gravitationally unbound substructure
to persist for much longer than a dynamical time. For a gravitationally bound protocluster, the dynamical
time $t_{\rm dyn}$ is of the same order as the free-fall collapse time of the protocluster, $t_{\rm ff}$,
and so our requirement for competitive accretion is the same as the classical requirement for efficient gravitational fragmentation, $t_{\rm cool} < t_{\rm ff}$ \citep{ReesOstriker1977}.

In local star-forming regions, we know that the cooling time is very short, and so this condition is easily
satisfied. Can the same thing be said for low-metallicity gas? In the limiting case of zero metallicity, we know from the many detailed simulations that have been performed that this is generally not the case. Owing to the relative ineffectiveness of $\mHt$ as a coolant, the gas cools slowly, and achieves at best a cooling time that is comparable to its dynamical time, but that is not significantly shorter (see e.g.\ Figure 6 in \citealt{OsheaNorman2007}). We therefore would not expect  competitive accretion to occur in this case, and indeed numerical simulations that start with proper cosmological initial conditions find little evidence for widespread fragmentation \citep[e.g.][]{Abeletal2002,Yoshidaetal2006}, although binary formation may occur in some cases (Turk et~al., 2009, in prep.). However, the addition of even a small quantity of metals to the gas 
 can dramatically alter this picture. The detailed chemical and thermodynamic evolution of low-metallicity gas has been the focus of several studies \citep{Schneideretal2002, Schneideretal2006, Omukaietal2005}, and the general features of their results are broadly similar (see, in particular, Fig.\ 1 of \citealt{Omukaietal2005} and Fig.\ 3 of \citealt{Schneideretal2006}). These studies have shown that there are two main phases during which metal-enriched gas can cool significantly, and on a timescale faster that the free-fall time. One occurs at low densities, and is caused by a combination of rotational and vibrational line cooling from H$_{2}$ and HD and atomic fine structure cooling from carbon, oxygen, silicon and iron. The other occurs at higher densities, and is caused by dust cooling. These two regimes are illustrated schematically in Fig. 1, and are discussed in turn in the following sections.

\begin{figure*}
\begin{center}
\includegraphics[width=5.0in]{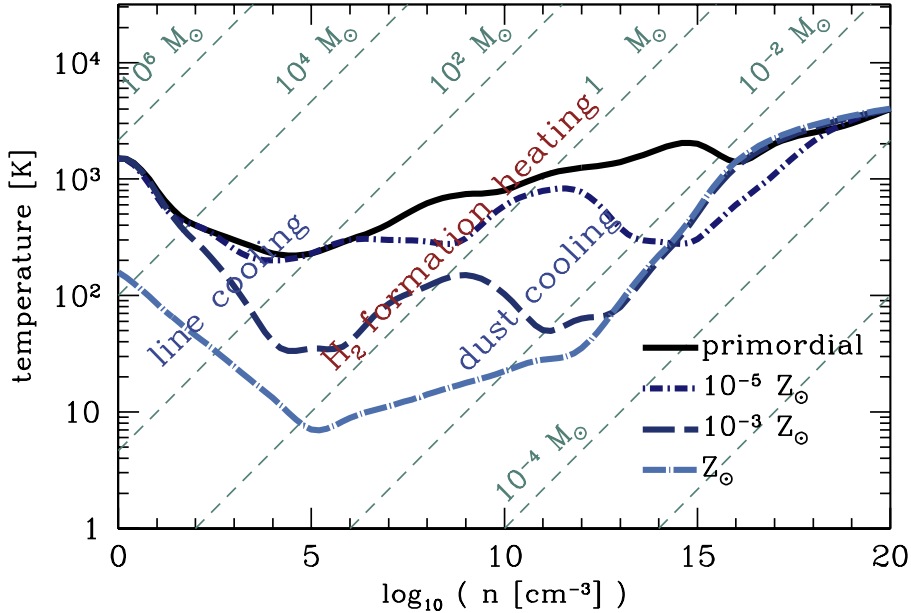}
\end{center}
\caption{\label{eosplot} We illustrate the expected equilibrium temperatures of the gas as a function of density for several levels of metal enrichment. Note that these are purely representative of the results from detailed studies. The solar metallicity line, for $n \la 10^{12} \rm cm^{-3}$, is an approximation of that proposed by \citet{Larson2005}, combined with the expected shift to an effective polytropic gamma of 1.4 at around  $n \approx 10^{12} \rm cm^{-3}$ once the gas become optically thick (e.g. \citealt*{Masunagaetal1998}), and then the subsequent cooling provided when the temperature is high enough to dissociate H$_{2}$. The other three curves for the lower metallicity and primordial gas are taken to follow (roughly) the results from the study by Omukai et al (2005), in particular their Fig. 1 (but see also \citealt{Schneideretal2002, Schneideretal2006}). All the metal enriched curves show two clear regimes: one in which the cooling is dominated by the fine-structure line-emission (for 
 $n \la 10^{6} \rm cm^{-3}$, depending on the metallicity), and a second in which the cooling is dominated by the dust. For lower metallicity gas the dominant formation process for H$_{2}$ is via three-body gas phase reactions, in contrast to the solar metallicity case in which the dust fraction is large enough for the grain surface reactions to dominate. The energy released during the three-body reactions offsets the cooling that the  H$_{2}$ provides, and as such the gas is unable to rid itself of the compressional heating caused by the gravitational collapse, resulting in the temperature rise between the two cooling regimes. Fragmentation has been shown to occur in the regimes where the temperature decreases with increasing density. }
\end{figure*}

\subsection{Low density cooling from line emission}
A number of studies, beginning with the work of \citet{Brommetal2001}, have examined the cooling of low metallicity gas caused by atomic and molecular line emission at densities 
$n < 10^{6} \: {\rm cm^{-3}}$ \citep{BrommLoeb2003, SantoroShull2006, Jappsenetal2007, Jappsenetal2009a, Jappsenetal2009b, SmithSigurdsson2007, Smithetal2009}. These studies have shown that fine structure line emission from various metals, notably carbon and oxygen, is able to efficiently cool low density, metal-enriched gas, provided that the degree of enrichment is large enough. A common way to quantify the required degree of enrichment is in terms of a critical metallicity ${\rm Z_{crit}}$. This is, loosely, the lowest metallicity for which metal line cooling becomes an effective coolant at some point during the gravitational collapse of the gas. For instance, if we ask what metallicity is required in order for metal line cooling to balance compressional heating, we obtain an answer than depends on both the temperature and density of the gas, since the line cooling rates and the compressional heating rate are functions of both of these variables. However, the gas does not 
 evolve arbitrarily in temperature-density space, and along any realistic trajectory, ${\rm Z_{crit}}$ always has a well-defined minimum that falls in the range $10^{-4} {\rm Z_{\odot}}  < {\rm Z_{crit}} < 10^{-3.5} \: {\rm Z_{\odot}}$ \citep{SantoroShull2006,Frebeletal2007,Shull2008}. 

Numerical simulations have confirmed that gas with a metallicity ${\rm Z > Z_{crit}}$ can cool efficiently and quickly at low densities until the temperature hits the floor set by the CMB temperature, $T_{\rm CMB} = 2.726 (1+z)$ \citep{Brommetal2001, SmithSigurdsson2007, Smithetal2009, Jappsenetal2009a, Jappsenetal2009b}. The gas is therefore able to increase the number of Jeans masses available for gravitational fragmentation, while preserving the substructure required to seed that fragmentation. We therefore potentially have ideal conditions for competitive accretion. However, in practice, there is a complication: if the initial gas mass in the cooling region was significantly less than a Jeans mass (as may be the case if the gas is initially confined by a gravitational potential dominated by dark matter, rather than by its own self-gravity), then the gas may not be able to cool to a low enough temperature to produce multiple Jeans masses, and hence may not fragment. In local star-forming regions, this is not generally a concern, as the gas temperature can reach very small values. However, at high redshift, the temperature floor set by the CMB is much higher, and hence this is far more of a concern. 

The various numerical studies that have been performed of metal-enriched gas evolving in early protogalaxies tend to support this concern. Simulations that start with idealized initial conditions
\citep{Brommetal2001,Jappsenetal2009a} produce large masses of cooling gas that can easily become multiply Jeans unstable and that readily fragment. On the other hand, simulations that follow the build-up of a metal-enriched protogalaxy from more realistic cosmological initial conditions tend to produce a much smaller mass of cooling gas \citep{SmithSigurdsson2007,Smithetal2009}. In these simulations, fragmentation occurs only for metallicities in the range ${\rm 10^{-3.5} \: Z_{\odot} < Z < 10^{-2.5} \: Z_{\odot}}$.  At lower metallicities, metal-line cooling never becomes important, and hence $t_{\rm cool}$ remains comparable to or larger than $t_{\rm ff}$ throughout. On the other hand, at metallicities ${\rm Z > 10^{-2.5} \: Z_{\odot}}$ the gas cools {\em too} quickly: it reaches the CMB temperature floor at an early stage in the collapse, while its Jeans mass is still comparable to the amount of mass contained within the cooling region. Thereafter, it evolves isothermally
  at $T_{\rm CMB}$, and hence cannot further increase the number of Jeans masses available for fragmentation. Similar results were found by \citet{Jappsenetal2009b} in a study of ${\rm Z = 10^{-3} \: Z_{\odot}}$ gas cooling from an initially ionized state: the enhanced ionization promoted rapid cooling, and allowed the gas to reach the CMB temperature floor while still at low density, the number of Jeans masses available for fragmentation at the transition to isothermal evolution was small, and so the gas did not fragment. 

The important role played by the CMB in regulating low density fragmentation was further highlighted by a numerical experiment performed by \citet{Smithetal2009}. They ran a simulation with ${\rm Z = 10^{-2} \: Z_{\odot}}$ in which the thermodynamical effects of the CMB were not included (i.e.\ the CMB temperature was artificially set to zero). In this case, the gas was able to cool down to around 20~K (compared to 70~K in the case with the CMB, as the simulated collapse occurred at a redshift $z=25$), allowing the number of Jeans masses of gas contained within the cooling region to be increased by a further factor of ten. This resulted in fragmentation, as one would expect, with a larger number of fragments forming in this case than in any of their other simulations.

The low density, line-dominated cooling regime does not extend far beyond $n = 10^{6} \: {\rm cm^{-3}}$. Above this density, all of the major fine structure coolants have reached local thermodynamic equilibrium (LTE), and hence the cooling time due to these processes tends to a constant value, while the free-fall time continues to decrease with increasing density. We therefore soon reach a regime in which $t_{\rm cool} > t_{\rm ff}$, at which point compressional heating starts to warm up the gas. In addition, H$_2$ formation heating also begins to make a significant contribution at around this point \citep{Omukaietal2005}. The cloud therefore enters a `loitering' phase,  allowing material to collect at these densities.

\subsection{High density cooling from dust continuum emission}
\label{sec:dust}
The loitering phase that begins once metal-line cooling can no longer cool the gas effectively continues until the onset of the second phase of efficient cooling. This occurs once the timescale for thermal energy transfer from the gas to the dust grains becomes small. To determine when this occurs, we need to know how the properties of the dust -- specifically, the grain size distribution and the ratio of dust mass to gas-phase metal mass -- compare to those of the dust present in local star-forming regions. The rate at which energy is transfered from gas to dust is proportional to the total available surface area of grains per unit volume of gas, $A_{\rm tot}$. We can write the ratio of $A_{\rm tot}$ in gas with metallicity 
${\rm Z}$ to $A_{\rm tot}$ in solar metallicity gas as
\begin{equation}
\frac{A_{\rm tot}({\rm Z})}{A_{\rm tot}({\rm Z_{\odot}})} = f_{\rm dust} \frac{{\rm Z}}{{\rm Z_{\odot}}}.
\end{equation}
If we assume that the dust in low-metallicity gas is simply a less abundant version of local dust (i.e.\ that it retains a \citet*{Mathisetal1977} size distribution and that the ratio of dust mass to gas-phase metal mass is the same as in local gas), then we would simply have $f_{\rm dust} = 1$, and in this case the onset of the second phase of efficient cooling begins at a density of roughly $n_{\rm cr} \sim 10^{9} ({\rm Z} / 10^{-3} \: {\rm Z_{\odot}})^{-2} \: {\rm cm^{-3}} $ \citep{Omukaietal2005}. In the more general case, where the dust need not have these properties, we can instead write the required density as $n_{\rm cr} \sim 10^{9} f_{\rm dust}^{-1}  ({\rm Z} / 10^{-3} \: {\rm Z_{\odot}})^{-2} \: {\rm cm^{-3}}$. The actual value of $f_{\rm dust}$ in high redshift, low metallicity systems is not known, and is an active field of research. The large quantities of dust observed in the host galaxies of $z \sim 6$ quasar host galaxies \citep{Bertoldietal2003, Maiolinoetal2004, Wangetal2008} suggest that some mechanism for efficiently producing large quantities of dust must be active in the early universe. On the other hand, the latest theoretical models of dust production in supernovae -- formerly the leading candidates for producing the observed high-redshift dust -- suggest that much of the dust that is initially formed in the supernova remnant is destroyed by the passage of the reverse shock, and that little survives to escape into the interstellar medium  \citep{Nozawaetal2007,BianchiSchneider2007}. It is therefore quite plausible that $f_{\rm dust} \ll 1$ in low metallicity gas, but equally it is also plausible that $f_{\rm dust} \sim 1$.

Once the transfer of energy from the gas to the grains becomes rapid, then the gas quickly cools down to the dust temperature $T_{\rm d}$. This sudden cooling occurs throughout the large quantity of gas accumulated during the loitering phase, making it particularly easy in this case to produce a highly Jeans unstable gas distribution. An example of this mechanism in action can be found in \citet*{Clarketal2008a}, where the temperature-density relations from \citet{Omukaietal2005} are approximated by a piecewise-barotropic equation of state. It was found in this study that vigorous fragmentation  occurs during the dip in the equation of state caused by the sudden onset of efficient dust cooling, resulting in a cluster in which  competitive accretion processes dominate. Indeed, the resulting mass function of the `sink particles' \citep{Bateetal1995} was consistent with a Salpeter slope for metallicities as low as \zfive.

\subsection{Implications for the IMF}
As the preceding sections show, there are two regimes in low-metallicity gas in which competitive accretion could possibly operate: a low-density regime in which the cooling is dominated by atomic and molecular line emission, and a higher density regime which is dominated by dust cooling. In local star forming regions, these two regimes almost overlap: molecular line cooling begins to become ineffective at densities $n \sim 10^{4} \: {\rm cm^{-3}}$ and temperatures $T \sim 10 \: {\rm K}$, owing to the large line opacities that have developed by this point, while dust cooling starts to becomes highly effective at only a slightly higher density
$n \sim 10^5 \: {\rm cm^{-3}}$ \citep{Goldsmith2001}.

In lower metallicity gas, these two regimes are more distinct. Their relative importance depends on two main factors: the metallicity of the gas, and the degree of heating provided by the cosmic microwave background or some other background heating source. At the lowest metallicities, neither metal-line cooling nor dust cooling is able to cool the gas rapidly enough to allow competitive accretion to occur, and the evolution of the gas takes place just as in the metal-free case \citep*{Clarketal2008a}.

At higher metallicities, dust and/or metal-line cooling become effective. The required metallicity depends on the properties of the dust. Dust cooling becomes effective at a metallicity ${\rm Z} \sim 10^{-5} f_{\rm dust}^{-1} \: {\rm Z_{\odot}}$, where $f_{\rm dust}$ quantifies the total available surface area of dust grains, as described in \S\ref{sec:dust} above, whereas metal-line cooling becomes effective for a critical gas-phase metallicity ${\rm Z_{\rm crit}} = 10^{-3.5} \: {\rm Z_{\odot}}$. Therefore, if $f_{\rm dust} \simless 0.03$, metal-line cooling becomes effective at a lower metallicity than dust cooling. In this regime, the gas cools faster than it can dynamically respond. However, it cannot cool indefinitely; eventually, it must reach a temperature at which further cooling is balanced by heating, either from compression, or from the effects of external sources of radiation or the CMB. The Jeans mass of the gas at the point where this phase of efficient cooling 
 finishes -- due, for instance, to the gas temperature reaching the CMB floor -- determines whether competitive accretion will be possible. It sets the characteristic mass scale for any protostellar cores that form, and since competitive accretion requires a collection of such cores, it requires that the mass of cooling gas available be much larger than the Jeans mass at this point. If this is the case, then the outcome will be an IMF with a power-law high-mass end, and a turnover corresponding to this characteristic mass scale. 

For example, if metal-line cooling is halted by the gas temperature reaching the CMB temperature, and this occurs at a gas density $n_{\rm CMB}$, then this characteristic mass scale can be written as \citep{Elmegreenetal2008}
\begin{equation}
M_{\rm char} = 10 \left(\frac{T_{\rm CMB}}{50 \: {\rm K}}\right)^{3/2} 
\left(\frac{n_{\rm CMB}}{10^{4} \: {\rm cm^{-3}}}\right)^{-1/2} \: {\rm M_{\odot}}.
\end{equation}
Typically, $n_{\rm CMB} \sim 10^{4} \: {\rm cm^{-3}}$ for metallicities close to ${\rm Z_{\rm crit}}$, and so at high redshift, we expect $M_{\rm char}$ in these systems to be much larger than a solar mass (c.f.\ the fragment mass distributions in \citealt{Jappsenetal2009a} and \citealt{Smithetal2009}).  If the metallicity is much larger than ${\rm Z_{\rm crit}}$, however, so that cooling is very efficient and $n_{\rm CMB}$ is small, then $M_{\rm char}$ may be comparable to the total mass of cooling gas, in which case gravitational fragmentation will be strongly suppressed and competitive accretion will not occur.  On the other hand, at low redshifts, $M_{\rm char} \sim 1 \: {\rm M_{\odot}}$ or less, and so in this case, we would expect a standard Salpeter IMF to be produced at all metallicities greater than the critical value.

If  $f_{\rm dust} \simgreat 0.03$, then dust cooling will become effective at a lower metallicity than 
metal-line cooling. In this case, rapid cooling, fragmentation and competitive accretion will take place at high densities, where the characteristic Jeans mass is low even if heating from the CMB keeps the dust temperature at a much higher level than is typical of local star-forming regions. Therefore, we would expect a standard Salpeter IMF extending down to relatively small masses to be produced in this scenario.

Once both dust {\em and} metal-line cooling have become important, the outcome depends on just how effective the metal-line cooling actually is. For metallicities ${\rm Z} \, \simless \, 0.01 \: {\rm Z_{\odot}}$, the metal-line cooling phase is followed by a heating phase prior to the onset of efficient dust cooling \citep{Omukaietal2005}, allowing the conditions for competitive accretion to be achieved for a second time during the same collapse, only now occurring in gas with a higher density, and hence lower characteristic mass. In this case, we would again expect a standard IMF. On the other hand, if  ${\rm Z} \simgreat 0.01 \: {\rm Z_{\odot}}$, then the metal-line cooling will keep the gas close to the CMB temperature until it becomes optically thick. In this case, even if dust cooling becomes effective at high densities, it cannot cause a significant drop in temperature; the gas continues to evolve isothermally, and competitive accretion does not occur. We would therefore expect to obtain a power-law IMF with $M_{\rm char} \gg 1 \: {\rm M_{\odot}}$ at high redshifts, smoothly transitioning to a standard IMF at low redshifts as $T_{\rm CMB}$ decreases.

Finally, it should be noted that we are not the first to suggest that there may be a link between the CMB temperature and $M_{\rm char}$ (see e.g.\ \citealt{Larson1998, Larson2005, ClarkeBromm2003,  Tumlinson2007}). However, previous work on this issue has generally focussed only on fragmentation during the metal-line cooling regime, and has concentrated on determining $M_{\rm char}$, rather than on determining the shape of the IMF at $M > M_{\rm char}$.

\section{Preventing competitive accretion}
\label{sec:prev} 

Given any cloud with multiple Jeans masses, in which the gas and the newly formed fragments are free to move in the potential, competitive accretion will occur. Essentially, the gas will always want to reduce its energy state by moving down the potential well. Preventing competitive accretion requires that the gas reservoir be held up against the gravitational force of the cluster, but, at the same time, {\em some} of the gas must be permitted  to form cores, or else no star formation would occur. In other words the ambient gas would need to behave as if it is `hot', while the cores are `cold'. In this section we discuss some of the processes that may lessen the effects of competitive accretion in real clusters.

\subsection{Supersonic turbulence}

Although most of the studies to have looked at competitive accretion have used calculations in which the supporting supersonic turbulent motions have been free to decay by dissipating energy in shocks, several complementary studies have looked at the case of driven turbulence. In particular, \citet{Klessen2001} examined how the turbulent driving scale affects the fragmentation of self-gravitating isothermal gas in multiply Jeans unstable clouds. It was found that when the turbulence was driven on scales larger than the Jeans length, the star formation occurred in clusters. In contrast, when the turbulence was driven on progressively smaller scales, star formation became a progressively more isolated event, with competitive accretion playing almost no role. In the clustered mode, the mass function was set by the competitive accretion which occurred within the cluster's potential, and was consistent with the field star IMF (see also \citealt{Jappsenetal2005}). In the later case of isolated star formation, competitive accretion played no role and the resulting mass function was considerably flatter than the observed distribution. A more general discussion of the role of supersonic turbulence in star formation can be found in \citet{MacLowKlessen2004, BallesterosParedesetal2007}

That competitive accretion can work under the conditions of large-scale driving in isothermal gas is not surprising: the ram pressure is not isotropic and entrains mass, such that the flows gather Jeans unstable material together to form large potential wells. Also, the turbulence is free to decay inside these over-densities, as it is typically shielded from the outside driving by a boundary shock. Once inside the potential, the gas is free to move under its influence when it has lost enough of its kinetic energy in shocks. This loss of kinetic energy occurs on the crossing time \citep{MacLowetal1998b, MacLowetal1998a} and, as discussed above, can help to promote fragmentation.  The fact that turbulence is observed to be dominated by large-scale modes supports this picture of cluster formation \citep{MacLowOssenkopf2000, Ossenkopfetal2001, OssenkopfMacLow2002,Brunt2003}. 

\subsection{Small-scale sources of turbulence}
In real clouds it is still not clear whether or not supersonic turbulence is also constantly driven on smaller scales within the cluster, in a way which could dilute the effects of competitive accretion -- putting aside for the moment that it appears to result in the wrong IMF (Klessen 2001). Jets and outflows from young protostars perhaps provide the best candidate for halting subsequent accretion (e.g. \citealt{Matzner2007}), but a full treatment of the problem has never been implemented. \citet{LiNakamura2006} and \citet{NakamuraLi2007} have looked at the effect of outflows (winds and jets) on the star formation rate in small clusters, concluding that they help it to self-regulate at a fairly low value. However these calculations do not attempt to model any subsequent accretion onto the `star' particles after their birth. Their numerical set-up also employs periodic boundary conditions, such that the outflow energy can never escape the cluster. In a different set-up, \citet{Banerjeeetal2007} performed high-resolution simulations of single jets, concluding that they were unable to generate space-filling supersonic turbulence. Rather they found that the turbulence was only supersonically driven in the bow-shock region, while the rest of the cloud exhibited sub-sonic turbulence.

Altogether, it is not clear from the current simulations that competitive accretion in such an environment would be halted, rather than just slowed down. The results of \citet{DaleBonnell2008}, which focus on the impact of winds from high mass stars, suggest the latter: the accretion/fragmentation process is slowed, but the winds have little effect on the mass function produced by competitive accretion. They find that both filaments and disks will very effectively collimate the outflows, allowing accretion to continue along the filaments. 

Ideally, jets and outflows have a better chance of kinetically heating the cluster gas reservoir -- and thus slowing down the competitive accretion process -- when there are multiple sources present in the cloud, since they will deposit their combined energy over a wider volume. For example, \citet{Banerjeeetal2007} point out that it would take $\sim 100$ jets to match the kinetic energy of a virialized cluster such as the Orion Nebula Cluster (ONC), assuming a perfect coupling between the jet and the cloud. If one assumes a constant rate at which stars are formed, $t_{\rm SF}$, and a constant lifetime of the outflow, $t_{\rm jet}$, then the number of outflow sources available at any one time is, $N_{\rm jet} \approx t_{\rm SF}/t_{\rm jet}$. For a system like the ONC with $\sim$1000 stars, and taking the typical lifetime of a jet to be $10^{5}$ yr, this implies a $t_{\rm SF}$ of less than $10^{6}$ yr,  faster than some timescales for the ONC in the literature (e.g. \citealt{Tanetal2006}).  In practice, a better measure of $t_{\rm jet}$ might be the time at which it leaves the main cluster-forming core, since it will have little influence after this point.  In addition, it has been suggested than the jet lifetime in the OMC1/2 is only a few $10^{4}$ yr \citep{Asoetal2000, Williamsetal2003}. If jets are to be important, then these shorter outflow lifetimes would suggest an even shorter cluster formation time. Finally, early results from self-consistent simulations of jets/outflows -- which include coupling between the measured accretion rate and the launch velocity -- suggest that the cloud is not substantially disrupted, and accretion is only marginally slowed \citep{Banerjeeetal2009a}.

As one goes to lower metallicity, the picture becomes even more uncertain. On the one hand, the winds from lower metallicity stars are expected to be weaker than those we see from solar metallicity star formation \citep{Kudritzki2002, VinkdeKoter2005}. On the other hand, it is likely that the thermodynamic behavior at lower metallicities leads to a higher rate of star formation, as defined by the rate at which new stars form ($t_{\rm SF}$), and so their combined effect may be greater due to a better volume coverage. This is especially true if the fragmentation is dominated by the dust-coupling regime, since it typically occurs at higher densities than the line cooling, and hence is described by a smaller free-fall time. However, for the winds to be effective, they still have to counteract the large-scale infall of the gas that is coming in from lower densities, and destroy the filamentary structure that it will posses from the large-scale gravitational instabilities \citep{KlessenBurkert2000, Enriqueetal2007, Bonnelletal2008}. In contrast, if the fragmentation takes place primarily in the line-cooling regime, then we would have a similar situation as in solar-metallicity star formation.

\subsection{Radiative feedback from young stars}

To date, the majority of the simulations used to study competitive accretion have neglected the effects of the accretion luminosity that is released as the gas falls onto the protostellar surface. Given that this radiation can alter the Jeans mass in the surrounding gas (e.g. \citealt{Krumholz2006}), and thus change the fragmentation characteristics around the protostars, it is instructive to ask whether this process can prevent competitive accretion.

In the picture described in \S \ref{sec:compacc}, all stars start off with masses comparable to the mean Jeans mass in the cloud, and so competitive accretion begins as low-mass star formation. The recent calculations by \citet{Bate2009} show that the heating from this low-mass star formation phase is important on scales of 100-1000 AU, much shorter than either the mean Jeans length in the parent cloud or in nearby star-forming regions. While this cuts down the number of objects forming in disks (in comparison to \citealt{Bateetal2003}), the larger-scale fragmentation is unaffected, and so is the dynamical interaction of the stars in the potential.

However as some of the stars start to accrete more mass, the accretion luminosity will start to rise, particularly if the accretion rate is high. For example \citet*{Krumholzetal2007} show that for accretion rates between 10$^{-4}$ to 10$^{-3}$ \solmas yr$^{-1}$, the gas can be heated to as much as 50~K at distances of around 2000 AU. By this point in the competitive accretion picture however, the majority of the new fragmentation is taking place in dense filaments which are squeezed as they fall into the potential \citep*{Bonnelletal2008}. Not only are these filaments well shielded by the radiation from the central accreting sources, but they become gravitationally unstable at distances $\ga$ 8000 AU, still well outside the 50~K gas. Further, \citet{Daleetal2005} showed that these filaments are also shielded from any ionizing sources at the cluster center.

At lower metallicities, and in the case where the only period of competitive accretion occurs in the regime dominated by atomic fine structure 
cooling, a similar argument applies. Fragmentation in this case occurs at similar densities to those at which it occurs in present-day star-forming
clouds, and since the characteristic mass is comparable to or larger than the present-day value, the associated length scales will also be comparable
to or larger than those in present-day clouds. 

If fragmentation occurs in the regime dominated by dust cooling, however, then the situation is somewhat more complicated.  As one moves to lower metallicities, the coupling to the dust -- and thus the fragmentation -- occurs at progressively higher densities. The question is then whether the heating of the dust by the accretion luminosity is sufficient to remove the dip in the temperature-density curve (see Figure 1). As the dip is caused by the gas cooling to reach the dust temperature, the dip can be removed if the heating due to the accretion luminosity raises $T_{\rm dust}$ above $T_{\rm gas}$ before the onset of efficient thermal coupling between gas and dust. How easy this is to bring about depends on the gas temperature before the onset of efficient thermal coupling. If the gas temperature is high, as will generally be the case in low metallicity systems, then the dust must also be heated to a high temperature. Since the rate at which the dust can re-radiate energy sc
 ales as a high power of the dust temperature (typically $\Lambda_{\rm dust} \propto T_{\rm dust}^{5}$ -- $T_{\rm dust}^{6}$ for $T_{\rm dust} < 100 \: {\rm K}$), the energy input required in this case is very large, and is only likely to be achieved in gas very close to the accreting protostar. More plausibly, heating due to accretion may reduce the depth of the dip in the temperature-density curve 
slightly, thereby increasing the characteristic mass. However, if cooling and fragmentation are occuring at high densities, then the characteristic mass will remain relatively small.  If the metallicity is higher, and the gas temperature is lower, then increasing $T_{\rm dust}$ to $T_{\rm gas}$ requires less
energy input, but at the same time, the thermal coupling and fragmentation will occur at lower densities, further from the protostar, and hence in a region where there is far less heating.


\subsection{Magnetic fields}
\label{sec:mag}

Magnetic fields are ubiquitously observed in interstellar gas on all scales \citep{Crutcheretal2003,HeilesTroland2005}. However, their importance for star formation and for the morphology and evolution of molecular cloud cores remains controversial. A crucial parameter in this debate is the ratio between core mass and magnetic flux.  In supercritical cores, this ratio exceeds a critical value and collapse can proceed. In subcritical cores, magnetic fields provide  stability \citep{Spitzer1978,Mouschovias1991b,Mouschovias1991a} at least if we assume ideal magnetohydrodynamics. Recent measurements of the Zeeman splitting of molecular lines in nearby cloud cores indicate mass-to-flux ratios that lie above the critical value, in some cases only by a small margin but very often by factors of many if non-detections are included \citep{Crutcher1999, CrutcherTroland2000, Bourkeetal2001, Crutcheretal2008}. 
 
We now consider the conditions for competitive accretion, as discussed in Section \ref{sec:fragtime}, in the context of supercritical and subcritical regimes. For subcritical cloud cores to form stars \citep{Shuetal1987}, ambipolar diffusion is required, i.e.\ the drift between neutral and charged particles. It causes a redistribution of the magnetic flux until the inner regions of the core become supercritical and go into dynamical collapse \citep{Mouschovias1976, Mouschovias1979, MouschoviasPaleologou1981}. This process was originally thought to be slow, because in highly subcritical clouds the ambipolar diffusion timescale, $t_{\rm AD}$, is about 10 times larger than the dynamical time, $t_{\rm ff}$. Regardless of the exact value of the ambipolar diffusion timescale,  in a magnetically dominated region the first two system timescales ($\tau_{\rm form}$, and $\tau_{\rm acc}$) are roughly equivalent to $t_{\rm AD}$. Their relative values remain unchanged and the first relation in condition \ref{equ:timescales} still holds. However, the last relation may break down for $t_{\rm AD} \gg t_{\rm ff}$, and feedback may remove the cluster gas before contracting cores have had time to interact or undergo further accretion from the common reservoir. 

However for cores close to the critical value, as is suggested by observations, both timescales are comparable. Numerical simulations furthermore indicate that the ambipolar diffusion timescale becomes significantly shorter for turbulent velocities similar to the values observed in nearby star-forming region  \citep{FatuzzoAdams2002,Heitschetal2004,LiNakamura2004}. In this case we fully recover both relations in Expression \ref{equ:timescales}, and the star forming regions should be subject to competitive accretion. If most cloud cores are magnetically supercritical anyway, then the relevant timescale is the dynamical one, $t_{\rm ff}$, and the discussion in Section \ref{sec:fragtime} is applicable. 

However, we note several points of caution. Although  magnetic fields in general appear to be too weak to prevent gravitational collapse to occur, they still may influence the star formation process in various ways. Recent numerical simulations have shown that even a weak magnetic field can  change the coupling between stellar feedback processes and their parent clouds \citep{NakamuraLi2007, Krumholzetal2007b}, or slow down the overall evolution \citep{Heitschetal2001} by a factor of two or so. Although this increases $\tau_{\rm acc}$, it will,  under normal conditions, remain smaller than $\tau_{\rm emerge}$. Competitive accretion still influences the evolution but at a lower level than in the purely hydrodynamic case. We also mention, that magnetic fields when considered in the ideal MHD approximation are able to reduce the fragmentation of cloud cores on very small scales \citep{PriceBate2007a, PriceBate2008, HennebelleFromang2008, HennebelleTeyssier2008} and influence the properties of protostellar disks  \citep{PriceBate2007b, MellonLi2008b, MellonLi2008a}. Due to magnetic breaking, disks tend to be smaller and less prone to gravitational instability. This efficiently suppresses binary formation in such models, and again points towards the necessity to consider non-ideal MHD (such as ambipolar diffusion  or Ohmic dissipation) in order to explain the observed binary distribution  \citep{Lada2006}. Our current analysis is only weakly affected by these uncertainties, because competitive accretion acts on scales much larger than individual protostellar disks.

\section{Summary}
\label{sec:summary}

Current studies of resolved stellar populations find the that high-mass end of the initial mass function (IMF) is well described by a Salpeter-like slope (Salpeter 1955). In this paper, we predict that whenever star formation occurs in bound stellar clusters, the slope of the mass function, above some characteristic mass, will be consistent with the Salpeter value. We suggest that this is due to the process of competitive accretion (e.g. Bonnell \& Bate 2006). 

In our discussion, we point out that competitive accretion is robust against both the protocluster density profile, and the details of the fragmentation within the protocluster's gravitational potential. 
The only requirement is that fragmentation can result in the formation of a bound group of protostars. 

We explore how fragmentation is likely to vary as a function of metallicity and show that for non-zero metallicities, there exist the necessary regimes where gas cools while collapsing, such that a single Jeans mass cloud will evolve towards having numerous Jeans masses. Depending on the metallicity of the gas, this cooling can be provided by either line emission, occurring at `low' densities ($n \la 10^{7} \rm cm^{-3}$) or dust continuum emission, occurring at higher densities. At solar metallicities, the two regimes coincide. The result of this cooling is fragmentation of the collapsing gas cloud, and the formation of a bound group of stars.

In conclusion, we predict that the Salpeter slope is a ubiquitous outcome of clustered star formation, regardless of the metallicity of the system.

\begin{acknowledgments}
The work carried out in this paper was supported by the European Commission FP6 Marie Curie RTN CONSTELLATION (MRTN-CT-2006-035890) and by a FRONTIER grant of Heidelberg University sponsored by the German Excellence Initiative. P.C.C. and R.S.K. acknowledge support from the German Science Foundation (DFG) under the Emmy Noether grant KL1358/1 and the Priority Program SFB 439 Galaxies in the Early Universe. S.C.O.G. acknowledges funding from the DFG via grant KL1358/4.
\end{acknowledgments}

\bibliographystyle{apj}
\bibliography{/Users/pcc/bibfile/pccbib}

\end{document}